\documentclass[letterpaper,twocolumn,english,aps,prl,floatfix,showpacs,tightenlines]{revtex4}
\usepackage[T1]{fontenc}
\usepackage[cp1250]{inputenc}
\usepackage{amsmath}
\usepackage{amssymb}
\usepackage{graphicx}
\usepackage{esint}

\makeatletter

\pdfpageheight\paperheight
\pdfpagewidth\paperwidth

\newcommand{\lyxdot}{.}

\@ifundefined{textcolor}{}
{%
 \definecolor{BLACK}{gray}{0}
 \definecolor{WHITE}{gray}{1}
 \definecolor{RED}{rgb}{1,0,0}
 \definecolor{GREEN}{rgb}{0,1,0}
 \definecolor{BLUE}{rgb}{0,0,1}
 \definecolor{CYAN}{cmyk}{1,0,0,0}
 \definecolor{MAGENTA}{cmyk}{0,1,0,0}
 \definecolor{YELLOW}{cmyk}{0,0,1,0}
 }

\newcommand{\tr}{\mathrm{tr}}

\newcommand{\1}{\leavevmode{\rm 1\ifmmode\mkern  -4.8mu\else\kern -.3em\fi I}}

\usepackage{babel}
\usepackage{times}
\usepackage{hyperref}

\makeatother

\usepackage{babel}
\begin{document}

\title{Universal time-fluctuations in near-critical out-of-equilibrium quantum
dynamics}

\author{Lorenzo Campos Venuti and Paolo Zanardi}

\affiliation{Department of Physics and Astronomy and Center for Quantum Information
Science \& Technology, University of Southern California, Los Angeles,
CA 90089-0484, USA}
\begin{abstract}
Out of equilibrium quantum systems, on top of quantum fluctuations,
display complex temporal patterns. Such time fluctuations are generically
exponentially small in the system volume and can be therefore safely
ignored in most of the cases. However, if one consider small quench
experiments, time fluctuations can be greatly enhanced. We show that
time fluctuations may become stronger than other forms of equilibrium
quantum fluctuations if the quench is performed close to a critical
point. 
For sufficiently relevant operators the full distribution function
of dynamically evolving observable expectation values, becomes a universal
function uniquely characterized by the critical exponents and the
boundary conditions. At regular points of the phase diagram and for
non sufficiently relevant operators the distribution becomes Gaussian.
Our predictions are confirmed by an explicit calculation on the quantum
Ising model. 
\end{abstract}
\maketitle

\paragraph{Introduction}

Low temperature quantum matter at equilibrium organizes itself in
different phases separated by critical regions featuring enhanced
quantum fluctuations \cite{patashinskii_fluctuation_1979,sachdev_quantum_2011}.
This is due to the existence of competing interaction terms in the
system's Hamiltonian each of which strives to order the system according
different symmetry patterns. When the system is taken out-of-equilibrium
e.g., by a sudden change of the Hamiltonian parameters, on top of
these quantum fluctuations, temporal fluctuations are present as well.
In a series of papers \cite{campos_venuti_unitary_2010,campos_venuti_universality_2010,campos_venuti_exact_2011,campos_venuti_gaussian_2013}
we have shown that the full time-statistics of a dynamically evolving
expectation value ${\cal A}(t)$ of a quantum observable $A$ provides
a wealth of information about the equilibration properties of the
system, finite-size precursors of quantum criticality \cite{campos_venuti_universality_2010}
as well as a tool to single out quantum integrability in an operational
fashion \cite{campos_venuti_gaussian_2013}. 

Let us start by setting the general stage of our investigations. Consider
a quantum system driven out of equilibrium by an Hamiltonian $H$
and evolving unitarily according to the Schrödinger equation ($\rho\left(t\right)=e^{-itH}|\psi_{0}\rangle\langle\psi_{0}|e^{itH}$).
For definiteness we will focus on a many-body quantum system defined
on a $d$-dimensional lattice of volume $V$, and $A$ being a local
extensive observable. A first natural question is now: what is the
typical size of temporal fluctuations $\mathcal{A}(t)=\tr\left[A\rho\left(t\right)\right]$
? 
Reimann in \citep{reimann_foundation_2008, reimann_equilibration_2012}
has proven, assuming the non-resonant condition for the energy gaps,
the following bound on the temporal variance of $\mathcal{A}$ %
\footnote{We use here overline to indicate (infinite) time average which we
define as $\overline{f}=\lim_{T\to\infty}T^{-1}\int_{0}^{T}f(t)dt$.%
}: 
\begin{equation}
\Delta\mathcal{A}^{2}:=\overline{\left(\mathcal{A}\left(t\right)-\overline{\mathcal{A}}\right)^{2}}\le\mathrm{diam}\left(A\right)^{2}\tr\left(\overline{\rho}^{2}\right).\label{eq:reimann_bound}
\end{equation}
 where $\mathrm{diam}(A)$ is the diameter of the spectrum of $A,$
a measures of the strength of $A$. As shown in \citep{campos_venuti_gaussian_2013}
(see also {\cite{gambassi_large_2012,zangara_time_2013}}), for clustering
initial states%
\footnote{By clustering we mean here the property that, for operators $A,B$
localized around $x_{A},\, x_{B}$, the correlation $\langle AB\rangle$
decays sufficiently fast (i.e.~exponentially) with the distance $\left\Vert x_{A}-x_{B}\right\Vert $. %
}, $\tr\left(\overline{\rho}^{2}\right)\lesssim e^{-\alpha V}$ so
that the normalized fluctuations $\Delta\mathcal{A}/\overline{\mathcal{A}}$
are bounded by $\Delta\mathcal{A}/\overline{\mathcal{A}}\lesssim e^{-\alpha V}$.
This shows that in this general situation time fluctuations are practically
absent %
\footnote{ It is important to notice that these time fluctuations are extremely
small compared to the quantum fluctuations which, for clustering state
$|\psi\rangle$ are of the order of $\Delta A/\langle A\rangle\sim V^{-1/2}$
\citep{haag_local_1992,campos_venuti_unitary_2010}. %
} and one can safely replace dynamically evolving quantities $\mathcal{A}(t)$
with their averages, i.e.~$\mathcal{A}(t))\simeq\overline{\mathcal{A}}$
i.e., equilibration is achieved. 

However, there are at least two situations in which time fluctuations
are greatly enhanced and in some cases may become even stronger than
equilibrium ones. One possibility is to consider systems of non-interacting
particles. The bound (\ref{eq:reimann_bound}) does not apply in this
case and rather a {}``Gaussian equilibration'' scenario sets in
whereby time fluctuations are seen to scale as $\Delta\mathcal{A}/\overline{\mathcal{A}}\sim V^{-1/2}$
\citep{campos_venuti_gaussian_2013} (see also {\cite{cassidy_generalized_2011,gramsch_quenches_2012,he_single-particle_2013,zangara_time_2013}}).
This seems to be a precise prediction of the folklore according to
which free systems show poor equilibration.

Another possibility is offered by a small quench experiment. With
this we mean to tune the initial state to be the ground state of a
given Hamiltonian and then perform a small sudden change in the Hamiltonian
parameters. Intuitively, if the quench is sufficiently small only
relatively few quasi-particles get excited and contribute to the equilibration
process. This in turn results in poor equilibration property, i.e.~large
time fluctuations. Roughly speaking this is a region of parameters
for which $\alpha V\lesssim1$ and so the bound (\ref{eq:reimann_bound})
becomes ineffective. As shown in \citep{campos_venuti_universality_2010}
this situation can be used to locate precursor of criticality on small
systems by looking at dynamically evolving quantities.

In this Letter we will analyze this infinitesimal quench scenario
in detail and show that the full time-probability distributions of
(properly rescaled) expectation values of observables feature a novel
type of universality in the infinite-volume limit. For sufficiently
relevant operators the full distribution function become a universal
function uniquely characterized by the critical exponents and the
boundary conditions. Whereas, for non sufficiently relevant operators
or at regular point of the phase diagram the distribution becomes
Gaussian.

\paragraph{Observable dynamics for small quench}

Consider then the following small quench scenario. The system is prepared
in the ground state of the Hamiltonian $H_{0}$ for $t<0$. At time
$t=0$ one suddenly switches on a small perturbation $B$ such that
the evolution Hamiltonian becomes $H=H_{0}+\delta\lambda B$, with
$\delta\lambda$ a small parameter %
\footnote{The precise definition of {}``small'' is given in the Supplemental
Material%
}. Expanding ${\cal A}(t)$ up to first order in $\delta\lambda$ using
Dyson expansion and the spectral resolution $H_{0}=\sum_{n}E_{n}|n\rangle\langle n|$,
one gets 
\begin{equation}
{\cal A}(t)=\overline{A}+\delta\lambda\sum_{n>0}\left(Z_{n}e^{-it(E_{n}-E_{0})}+\mathrm{c.c.}\right)+O\left(\delta\lambda^{2}\right),\label{eq:full_expansion}
\end{equation}
where the first, time-independent term is the average of $\mathcal{A}\left(t\right)$
and with $Z_{n}:=A_{0,n}B_{n,0}/\left(E_{n}-E_{0}\right)$. The leading
contribution to the temporal variance is therefore at second order
and assuming that the gaps $E_{n}-E_{0}$ are non-degenerate one obtains
\begin{equation}
\Delta\mathcal{A}_{B}^{2}=2\delta\lambda^{2}\sum_{n>0}\left|Z_{n}\right|^{2}+O\left(\delta\lambda^{3}\right).\label{eq:var_AB}
\end{equation}
 We added a subscript $B$ to recall that the variance is computed
with perturbation $B$. Eq.~(\ref{eq:var_AB}) shows a intriguing
similarity to the zero temperature equilibrium isothermal susceptibility
defined by $\langle\psi\left(\delta\lambda\right)|A|\psi\left(\delta\lambda\right)\rangle=\langle\psi\left(0\right)|A|\psi\left(0\right)\rangle+\delta\lambda\chi_{AB}+O\left(\delta\lambda^{2}\right)$
($|\psi\left(\delta\lambda\right)\rangle$ being the ground state
of $H$). Indeed we can write 
$\chi_{AB}=2\sum_{n>0}\mathrm{Re}Z_{n}.$ 
 Inasmuch a super-extensive scaling of the susceptibility can be used
to detect criticality the same can be said for the time fluctuations.

Using Eq.~(\ref{eq:full_expansion}) we can actually obtain the full
probability distribution of the variable $\mathcal{A}$. Assuming
rational independence (RI) of the gaps $E_{n}-E_{0}$ and using the
theorem of averages we can express the time average as a phase space
average over a large dimensional torus. We then obtain, for the characteristic
function of $\mathcal{A}$, 
\begin{equation}
\overline{e^{is(\mathcal{A}-\overline{\mathcal{A}})/\delta\lambda}}=\prod_{n>0}J_{0}\left(2s\left|Z_{n}\right|\right):=J^{\mathcal{A}}\left(s\right),
\end{equation}
 where $J_{0}$ is the Bessel function of the first kind. So the probability
distribution of $\mathcal{A}$ is completely encoded in the function
$J^{\mathcal{A}}\left(s\right)$. Let us also define the Wick rotated
function $I^{\mathcal{A}}\left(s\right)=J^{\mathcal{A}}\left(is\right)$
which has the advantage of being positive for real $s$. We can then
define the coefficients $a_{n}$ by the series $\ln\left[I_{0}\left(s\right)\right]=\sum_{n=1}^{\infty}a_{n}s^{n}/n!$
which converges absolutely in a neighborhood of the origin. Note that
$a_{p}=0$ for $p$ odd. The cumulant of the variable $(\mathcal{A}-\overline{\mathcal{A}})/\delta\lambda$
are given by $\kappa_{2p}=a_{2p}2^{2p}Q_{2p}$ with $Q_{2p}:=\sum_{n>0}\left|Z_{n}\right|^{2p}$
(odd cumulants are zero). Under the assumption of convergence the
probability distribution of $\mathcal{A}$ is uniquely characterized
by the coefficients $Q_{2p}$. Conversely the probability distribution
uniquely defines the coefficients $Q_{2p}$ which are generalizations
of the variance Eq.~(\ref{eq:var_AB}). Intuitively, at critical
points the cumulants $\kappa_{2p}$ (through the coefficients $Q_{2p}$)
tend to diverge with the system size, higher cumulant being more divergent.

Let us analyze the behavior of $Q_{2p}$ close to quantum criticality.
In this case $\delta\lambda=\left|\lambda-\lambda_{c}\right|$ measures
the distance from the critical point $\lambda_{c}$. Using standard
scaling arguments one can show that $Q_{2p}\propto L^{2p\alpha}$
with $\alpha=2d+\zeta-\Delta_{A}-\Delta_{B}$ (see Supplemental Material).
Here $\Delta_{A/B}$ are the scaling dimensions of the observables
$A/B$ that we assumed extensive and $\zeta$ is the dynamical critical
exponent. Instead, away from criticality the expectation is $Q_{2p}\propto L^{d}$.
Requiring that, at finite size, $Q_{2p}$ is analytic in the system
parameters and matches the above scaling, one can predict the behavior
of $Q_{2p}$ close to the critical point both in the critical region
$\xi\gg L$ and in the off-critical one $\xi\ll L$: 
\begin{equation}
\kappa_{2p}\sim Q_{2p}\sim\begin{cases}
L^{2\alpha p} & \xi\gg L\\
\delta\lambda^{d\nu-2\alpha p\nu}L^{d} & \xi\ll L
\end{cases}\,.\label{eq:Q_scaling}
\end{equation}

Let us compare the strength of the temporal fluctuations encoded in
Eq.~(\ref{eq:Q_scaling}) with other familiar forms of quantum fluctuations
close to criticality. Equilibrium quantum fluctuations of an observable
$A$ in a state $|\psi\rangle$ are encoded in the cumulants $\langle A^{n}\rangle_{c}$
where the subscript $c$ denotes connected averages with respect to
$|\psi\rangle$. In the critical region the singular part of these
cumulants scales as $L^{n(d-\Delta_{A})}$ where $\Delta_{A}$ is
the scaling dimension of $A$ \citep{campos_venuti_unitary_2010}.
When one is interested in the response to a perturbation $A$ encoded
in a Hamiltonian $H=H_{0}+\lambda A$, other generalized cumulants
are given by the higher order susceptibilities $\partial^{n}E/\partial\lambda^{n}$
($E$ is the ground state of $H$, or the free energy at positive
temperature)%
\footnote{The term generalized cumulants can be understood given that these
quantities are derivatives of a --generalized-- cumulant generating
function, the free energy, i.e.~$\tr\exp-\beta(H_{0}+\lambda A)=\exp(-\beta F)=\exp(-\beta\sum_{n}\partial^{n}F/\partial\lambda^{n}\lambda^{n}/n!)$. %
}. At criticality such generalized susceptibilities grow as $L^{-\zeta+n/\nu}$,
in particular one as $L^{2d+\zeta-2\Delta_{A}}$ for $n=2$. Comparing
with Eq.~(\ref{eq:ratios}) we see that temporal fluctuations --which
are basically absent for general quenches-- become the strongest fluctuations
for small quenches close to criticality. Indeed, looking at the scaling
of the variances (and setting $A=B$ for simplicity) the exponents
for the temporal variance, susceptibilities and quantum variance satisfy
$2(2d+\zeta-2\Delta_{A})>(2d+\zeta-2\Delta_{A})>2d-2\Delta_{A}$.
As noted in \citep{campos_venuti_universality_2010} this opens up
the possibility of observing dynamical manifestations of criticality
on small systems.

Consider now the rescaled random variable $\mathcal{R}\left(t\right)=(\mathcal{A}\left(t\right)-\overline{\mathcal{A}})/\Delta\mathcal{A}$
whose cumulants are given by $\kappa_{2n}^{\mathcal{R}}=\kappa_{2n}^{\mathcal{A}}/(\kappa_{2}^{\mathcal{A}})^{n}$
for $n\ge1$ whereas odd cumulants are zero. The probability distribution
of $\mathcal{R}$ is uniquely determined by the ratios $R_{2p}=Q_{2p}/\left(Q_{2}\right)^{p}$.
From Eq.~(\ref{eq:Q_scaling}) we see that in the quasi-critical
regime, these ratios are scale independent and define some presumably
universal constants. Let us now find these constants. With the help
of density of states $\rho\left(E\right)=\tr\left(\delta\left(H-E\right)\right)$
we can write $Q_{p}=\int Q_{p}\left(E\right)\rho\left(E\right)dE$.
Since $\rho\left(E\right)dE$ is scale invariant, from $Q_{p}\propto L^{p\alpha}$
we derive $Q_{p}\left(E\right)\propto E^{-p\alpha/\zeta}$. In order
to proceed further we must assume the form of the low energy dispersion.
The simplest possibility is a rotationally invariant spectrum at small
momentum $E\simeq C\left\Vert \boldsymbol{k}\right\Vert ^{\zeta}=C(\sum_{j}k_{j}^{2})^{\zeta/2}$
where $\boldsymbol{k}$ is a quasi-momentum vector. In one dimension
this is essentially the only possibility but for $d>1$ one can also
have anisotropic transitions where the form of the dispersion depends
on the direction. Using the isotropic assumption we obtain $Q_{p}\simeq C'\sum_{\boldsymbol{k}}\left\Vert \boldsymbol{k}\right\Vert ^{-p\alpha}.$
In doing so we have essentially restricted the sum over $n$ to the
one-particle contribution. This is expected to be the leading contribution
whereas higher particle sectors contribute at most to the extensive,
regular term. At this point, a part from an irrelevant constant $C'$,
the behavior of the cumulants is \emph{uniquely} specified by the
critical exponent $\alpha$ and the \emph{boundary conditions} that
specify $\boldsymbol{k}$. More precisely the probability distribution
of the rescaled variable $\mathcal{R}\left(t\right)$ is a universal
function which depends only on $\alpha$ and the boundary conditions.
Let us be more explicit. Assume for concreteness that the lattice
is a hyper-cube of size $L$ and the boundary conditions (BC) are
such that moments are quantized according to $\boldsymbol{k}=(2\pi/L)(\boldsymbol{n}+\boldsymbol{b})$
with $n_{i}=1,\ldots,L$. The BC on the direction $i$ are fixed by
$b_{i}\in\left[0,1/2\right]$ which interpolate between periodic (PBC,
$b_{i}=0$) and anti-periodic (ABC, $b_{i}=1/2$) BC. Define now the
generalized $d$-dimensional Hurwitz-Epstein $\zeta$-function as
$\zeta_{\boldsymbol{b}}\left(\alpha\right)=\sum_{n_{1}=1}^{\infty}\cdots\sum_{n_{d}=1}^{\infty}\left\Vert \boldsymbol{n}+\boldsymbol{b}\right\Vert ^{-\alpha}$.
The cumulants of $\mathcal{R}$ depend on the universal ratios $R_{2p}=Q_{2p}/\left(Q_{2}\right)^{n}$
which are given by (letting $L$ go to infinity) $R_{2p}=\zeta_{\boldsymbol{b}}\left(2p\alpha\right)/\zeta_{\boldsymbol{b}}\left(2\alpha\right)^{p}$.
For example, in $d=1$ and for PBC one has $R_{2p}=\zeta\left(2p\alpha\right)/\zeta\left(2\alpha\right)^{p}$
where $\zeta\left(\alpha\right)$ is the familiar Riemann zeta function.
Clearly, anisotropic energy dispersion can also be treated introducing
even more general zeta functions with different exponents in different
directions. So far the exponent $\alpha$ has been quite arbitrary.
Indeed $\alpha$ can even become negative if $A$ and $B$ are not
sufficiently relevant. For example $\alpha=-\zeta$ if both $A$ and
$B$ are marginal operators. Apparently the moments of $\mathcal{R}$
become not normalizable in this case. The correct procedure is to
keep the sum over $\boldsymbol{k}$ finite, compute the leading finite
size correction, calculate the ratios $R_{2p}$ and then take the
infinite volume limit. The result is 
\begin{equation}
\lim_{L\to\infty}R_{2p}=\begin{cases}
\delta_{p,1} & 2\alpha\le d\\
\zeta_{\boldsymbol{b}}\left(2p\alpha\right)/\zeta_{\boldsymbol{b}}\left(2\alpha\right)^{p} & 2\alpha>d.
\end{cases}\label{eq:ratios}
\end{equation}
For $2\alpha\le d$ the characteristic function of $\mathcal{R}(t)$
becomes $e^{-s^{2}/2}$ in the thermodynamic limit and so $\mathcal{R}$
tends in distribution to Gaussian. Clearly the Gaussian behavior observed
here for not sufficiently relevant operators, is also to be expected
at regular points of the phase diagram. A discussion of the regular
points as well as a comparison of the dynamical central limit type
theorem here discussed and the one for quantum fluctuations at equilibrium
can be found in the Supplemental Material.

\paragraph{Loschmidt echo}

\begin{figure}
\begin{centering}
\includegraphics[width=6.6cm]{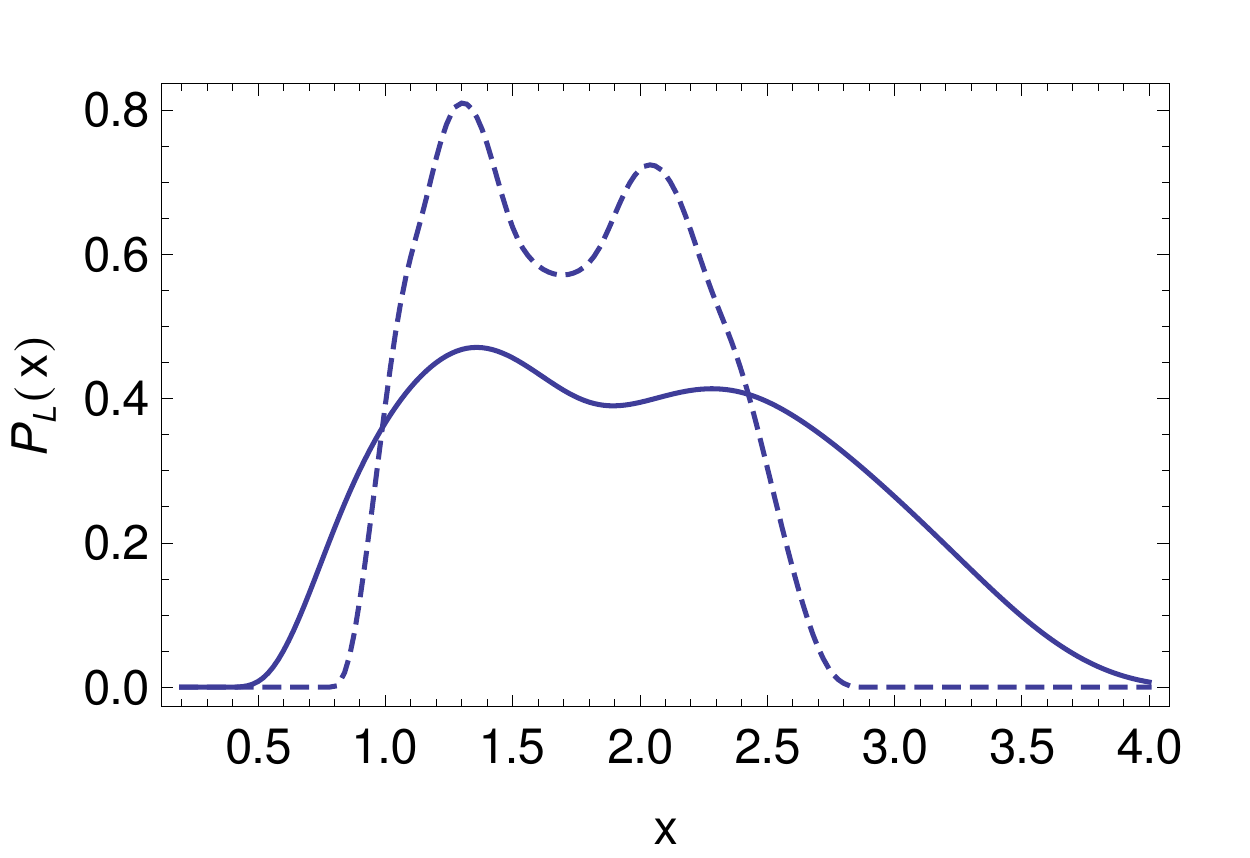} 
\par\end{centering}

\noindent \begin{centering}
\includegraphics[width=6.4cm,height=4cm]{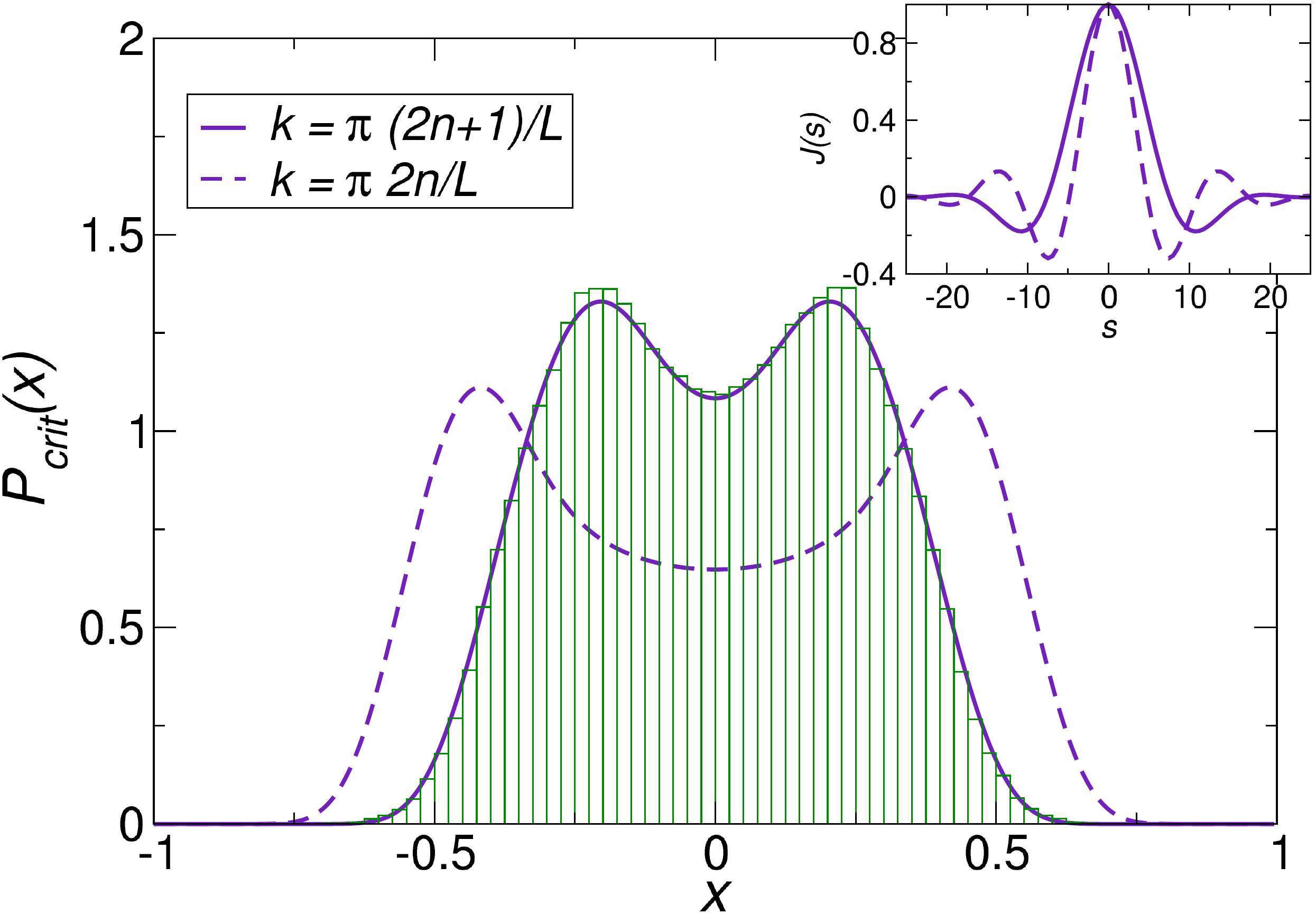} 
\par\end{centering}

\caption{\emph{Upper panel}: Universal probability distribution for the Loschmidt
echo in one dimension for different $\nu$ and PBC momenta i.e.~$k=2\pi n/L$.
Solid curve $\nu=1$ ($\alpha=2$) dashed $\nu=4/3$ ($\alpha=3/2$).
The curves are obtained integrating numerically Eq.~(\ref{eq:pdf}).
\emph{Lower panel}: Critical probability distribution for the quantity
$[\langle M(t)\rangle-\overline{M}]/(L\delta h)$ close to the Ising
critical point $h=1$. The inset shows the characteristic function
$J\left(s\right)=\prod_{n=0}^{\infty}J_{0}\left(s/\left(2n+1\right)\right)$.
The dashed lines are obtained using quasimomenta of the form $k=2\pi n/L$.
The histogram is computed performing a numerical experiment on the
Ising model on a chain of $L=1006$ sites with periodic boundary conditions.
The quench parameters are $h_{1}=1$, $h_{2}=1.0003$ and $\gamma_{1}=\gamma_{2}=1$.
The statistics is obtained sampling 600,000 random times uniformly
sampled in $\left[0,T\right]$ with $T=600,000$. The distribution
is unchanged using a different $\gamma_{1}=\gamma_{2}\neq0$ as implied
by universality.\label{fig:LE-critical}}
\end{figure}

Let us now extend the formalism by considering a particular, non-extensive
observable $A=|\psi_{0}\rangle\langle\psi_{0}|.$ In this case ${\cal A}(t)$
becomes the so-called the Loschmidt echo (LE) or survival probability
given by $\mathcal{L}\left(t\right)=\left|\langle\psi_{0}|e^{-itH}|\psi_{0}\rangle\right|^{2}$.
The Loschmidt echo is essentially the Fourier transform of the work
distribution function and is currently at the center of much theoretical
work. We will show that it is possible to obtain its full time statistics
exactly for a general initial state. Using the spectral resolution
of $H$ the LE can be written as $\mathcal{L}\left(t\right)=\left|G\left(t\right)\right|^{2}$
with $G\left(t\right)=\sum_{n}p_{n}e^{-itE_{n}}:=X\left(t\right)+iY\left(t\right)$
where $p_{n}=\left|\langle n|\psi_{0}\rangle\right|^{2}$ and $X(t),\, Y(t)$
are its real and imaginary part. Let us start by noticing that
\begin{equation}
\mathrm{Prob}\left(\mathcal{L}<r\right)=\int_{x^{2}+y^{2}<r}P_{X,Y}\left(x,y\right)dxdy\label{eq:prob_joint}
\end{equation}
where $P_{X,Y}\left(x,y\right)$ is the joint probability density
of $X$ \emph{and $Y$}, i.e. $P_{X,Y}\left(x,y\right)=\overline{\delta\left(X-x\right)\delta\left(Y-y\right)}$.
The related, joint characteristic function $\chi\left(\xi,\eta\right)=\overline{e^{i\xi X}e^{i\eta Y}}$
can again be computed as a phase-space average assuming rational independence
of the energies $E_{n}$. Expressing Eq.~(\ref{eq:prob_joint}) in
terms of $\chi\left(\xi,\eta\right)$, integrating over $x$ and $y$
and differentiating with respect to $r$ we obtain the following expression
for the probability density of the Loschmidt echo (more details in
the Supplemental Material) 
\begin{eqnarray}
P_{\mathcal{L}}\left(x\right) & = & \int_{0}^{\infty}K\left(x,\rho\right)J^{\mathcal{L}}\left(\rho\right)d\rho\label{eq:pdf}\\
K\left(x,\rho\right) & = & \frac{J_{1}\left(\sqrt{x}\rho\right)}{2\sqrt{x}}+\frac{\rho}{4}\left[J_{0}\left(\sqrt{x}\rho\right)-J_{2}\left(\sqrt{x}\rho\right)\right]
\end{eqnarray}
and $J^{\mathcal{L}}\left(\rho\right):=\prod_{n}J_{0}\left(p_{n}\rho\right).$
The probability distribution of the LE is completely encoded in the
function $J^{\mathcal{L}}$. Proceeding as previously we realize that
$J^{\mathcal{L}}\left(\rho\right)=\exp\left[\sum_{n}a_{n}\tr\left(\overline{\rho}^{n}\right)\left(-i\rho\right)^{n}/n!\right]$
where $\overline{\rho}$ is the infinite time average of $\rho\left(t\right)$.
Again, under assumption of convergence, the probability distribution
of \emph{$\mathcal{L}$ }is uniquely specified by the numbers $\tr\left(\overline{\rho}^{n}\right)$
and vice-versa. Since $\overline{\rho}$ is a positive operator this
uniquely specifies the spectrum of $\overline{\rho}$. Note that the
average LE is given by the first term in the expansion of $J^{\mathcal{L}}\left(\rho\right)$
and is $\overline{\mathcal{L}}=\tr\left(\overline{\rho}^{2}\right)$.
Let us first investigate the case of small quench close to a critical
point. It has been shown already that, as a function of energy the
weights $p\left(E_{n}\right):=p_{n}$ behave as $p\left(E\right)\sim\delta\lambda^{2}E^{-2/(\zeta\nu)}$
in the quasi-critical regime \citep{campos_venuti_universality_2010,de_grandi_quench_2010}.
Using again the isotropic assumption $E\sim\left\Vert \boldsymbol{k}\right\Vert ^{\zeta}$
we find $p_{\boldsymbol{k}}\sim\delta\lambda^{2}\left\Vert \boldsymbol{k}\right\Vert ^{-2/\nu}$.
Let us now study the rescaled LE $\mathcal{L}\left(t\right)/\overline{\mathcal{L}}$.
Its distribution is determined by $J^{\mathcal{L}}(\rho/\sqrt{\overline{\mathcal{L}}})=\exp\left[\sum_{n=1}a_{2n}T_{2n}(-\rho^{2})^{n}/(2n)!\right]$
with $T_{2n}:=\tr\left(\overline{\rho}^{2n}\right)/\tr\left(\overline{\rho}^{2}\right)^{n}$.
Once again, at criticality the probability distribution is uniquely
determined by the scaling exponent $\nu$ and the quantization of
the quasi-momenta $\boldsymbol{k}$. Moreover, taking the thermodynamic
limit ($L\to\infty$) in the quasi-critical region, the rescaled cumulants
$T_{2n}$ become exactly $T_{2n}=R_{2n}$ of Eq.~(\ref{eq:ratios})
with $\alpha=2/\nu$. In particular we see that for $4\nu\le d$ $\lim_{L\to\infty}J^{\mathcal{L}}(\rho/\sqrt{\overline{\mathcal{L}}})=\exp\left[-\rho^{2}/4\right]$.
Using this expression and Eq.~(\ref{eq:pdf}) one gets the probability
distribution of the rescaled LE: $\lim_{L\to\infty}P_{\mathcal{L}/\overline{\mathcal{L}}}\left(x\right)=\vartheta\left(x\right)e^{-x}$
i.e.~a Poissonian distribution. This in turns implies, for the un-rescaled
variable, $P_{\mathcal{L}}\left(x\right)\simeq\vartheta\left(x\right)e^{-x/\overline{\mathcal{L}}}/\overline{\mathcal{L}}$
a result which has been observed in \citep{campos_venuti_unitary_2010}
(see also the recent \citep{ikeda_emergent_2013}). Clearly the Gaussian
behavior of $J^{\mathcal{L}}$ predicted for not sufficiently relevant
operators $\nu\le d/4$ is expected to be seen also in the off-critical
region. In fact, the same considerations regarding the $\mathcal{R}$
variable apply in this case. A small quench in the vicinity but not
at the critical point has the effect of opening up a mass gap. This
in tun cures the infra-red divergences (UV divergences are cured by
the lattice) and one obtains that $\lim_{L\to\infty}T_{2n}=\delta_{n,1}$.
One also expects that this behavior extends a fortiori for more general
quenches. Indeed we conjecture that $\lim_{L\to\infty}T_{2n}=\delta_{n,1}$
for any sufficiently clustering initial state $|\psi_{0}\rangle$.
A plot of the universal, critical distribution of the LE is shown
in Fig.~\ref{fig:LE-critical}.

Computing critical distribution functions is a very hard task at equilibrium
as one needs the exact analytic form of the characteristic function.
For this reasons results are essentially limited to models that can
be mapped to a system of non-interacting particles such as the 2D
classical Ising model and its 1D quantum counterpart (see e.g.~\citep{lamacraft_order_2008}).
The situation for the temporal fluctuation is analogous and in order
to exemplify the formalism we consider now the 1D quantum XY model
with PBC. We perform a small quench in the transverse field close
to the Ising critical point at $h=1$. The expectation value of the
total magnetization takes the form $\mathcal{M}\left(t\right)=\sum_{j}\langle\sigma_{j}^{z}(t)\rangle=\overline{\mathcal{M}}+\sum_{k}\sin(\vartheta_{k}^{(2)})\sin(\delta\vartheta_{k})\cos(t\Lambda_{k})$.
The function $W_{k}:=\sin(\vartheta_{k}^{(2)})\sin(\delta\vartheta_{k})$
plays the role of $Z_{n}$ in Eq.~(\ref{eq:var_AB}). The characteristic
function can be computed assuming rational independence of the one
particle energies and one obtains
\begin{equation}
\overline{e^{i\lambda(\mathcal{M}-\overline{\mathcal{M}})/\Delta\mathcal{M}}}=\exp\sum_{k}\ln\left[J_{0}\left(\lambda W_{k}/\Delta\mathcal{M}\right)\right].\label{eq:chi-M}
\end{equation}
The scaling dimensions are in this case $d=\zeta=\Delta_{A}=\Delta_{B}=1$
implying that $\sum_{n>0}\left|Z_{n}\right|^{2}\sim L^{2}$. This
can be in fact proven analytically as, for small quench, $W_{k}$
is singular at $k=\pi$, where it behaves as $W_{k}\sim1/\left|\gamma\left(k-\pi\right)\right|$.
Expanding the argument of the exponential in series one realizes that
only the divergent part of $W_{k}$ is needed when computing the limit
$L\to\infty$. Given the fact that, for large $L$, $\Delta\mathcal{M}^{2}=2^{-1}\left(L/2\pi\right)^{2}\sum_{n=0}^{\infty}\left(n+1/2\right)^{-2}$,
we then obtain 
$\lim_{L\to\infty}\sum_{k}\ln\left[J_{0}\left(\lambda W_{k}/\Delta\mathcal{M}\right)\right]=\sum_{n=0}^{\infty}\ln J_{0}(\lambda\alpha_{n})$,
where $\alpha_{n}^{-1}:=\sqrt{\zeta_{1/2}\left(2\right)/2}\,(n+1/2)$%
\footnote{ Note that the equilibrium analogue of the above distribution $P_{\mathrm{eq}}(x)=\langle\delta\left(M-x)\right)\rangle$
instead is Gaussian for large size because the $M$ is not sufficiently
relevant ($2(d-\Delta_{M})=0\ngtr d=1$) as shown in \citep{eisler_magnetization_2003}.%
}. In figure \ref{fig:LE-critical} we plot the exact, critical, probability
distribution of the transverse magnetization obtained Fourier transforming
numerically Eq.~(\ref{eq:chi-M}). We also show very good agreement
with a numerical, small quench experiment performed on a XY chain.
The critical distribution is observed as long as $\xi\simeq\delta h^{-1}\gg L$
and $L\gtrsim20$. For shorter sizes, $\mathcal{M}$ is a sum of few
random variables and can be well approximated by retaining the two
dominant variables \citep{campos_venuti_universality_2010}. In the
off-critical region $\xi\ll L$ one obtains a Gaussian distribution
\citep{campos_venuti_gaussian_2013}.

\paragraph{Applications}

We would like now to point out the possible use of the time-fluctuations
formalism to distinguish critical or gapped regions in non-homogeneous
systems. A concrete realization of these systems is offered by optical
lattices of cold atoms in harmonic traps. Traditionally \cite{batrouni_mott_2002,rigol_local_2003,wessel_quantum_2004,rigol_numerical_2004}
Mott-insulating regions are distinguished from superfluid ones by
looking either at the quantum fluctuations of the particle densities
$\langle n_{i}^{2}\rangle-\langle n_{i}\rangle^{2}$ or at the local
compressibility (or suitably averaged version thereof) $\partial\langle n_{i}\rangle/\partial\mu_{i}$,
i.e.~a susceptibility. Small (resp.~large) fluctuations correspond
to Mott-insulating, {}``gapped'' (resp.~superfluid, {}``critical'')
regions. Indeed, as expected by the larger scaling and confirmed in
\citep{rigol_local_2003} the local compressibility is so far the
best indicator of insulating/superfluid region. The study of temporal
fluctuations for small quenches in trapped cold atom systems, may
offer an experimentally accessible yet powerful way to investigate
such non-homogeneous systems. The feasibility of such an approach
is currently under investigation \citep{yeshwanth__????}.

\paragraph{Conclusions}

In this Letter we have shown that the temporal fluctuations of quantum
observables for a small Hamiltonian quench near a critical poins feature
a novel type of universality that mirrors the one of quantum fluctuations
at equilibrium. The initial quantum state is choesen to be the ground
state of a Hamiltonian $H_{0}$ which is then slightly perturbed to
$H=H_{0}+\delta\lambda B$. 
Given the observable $A$ the temporal probability distribution $P_{\mathrm{dyn}}(a):=\overline{\delta\left(\langle A(t)\rangle-a\right)}$
(overline denotes the infinite time average) becomes Gaussian for
regular points of the phase diagram whereas it acquires a universal
form at critical points. Assuming hyperscaling, the critical distribution
function $P_{\mathrm{dyn}}$ is uniquely characterized by the critical
exponents and the boundary conditions it is hence even more universal
than the the equilibrium case. Moreover universal dynamical distributions
are observed even for less relevant operators. A byproduct of this
analysis is that, in the critical regime, temporal fluctuations are
stronger than other forms of equilibrium quantum fluctuations. This
opens up the possibility of assessing the critical character of non-homogeneous
systems by performing quench experiments.


The authors wish to thank H.~Saleur for precious discussions. This
research was partially supported by the ARO MURI under grant W911NF-11-1-0268.
PZ also acknowledges partial support by NSF grants No.~PHY-969969.

\bibliographystyle{unsrt}
\bibliography{temporal_transitions}

\section*{Supplemental Material}

\subsection{Small quench regime and universality}

The small quench regime can be encoded by the relation ~$\tr\left(\overline{\rho}^{2}\right)\simeq1.$
If this condition is met the bound (\ref{eq:reimann_bound}) becomes
ineffective. For small quench $\tr\left(\overline{\rho}^{2}\right)$
can be related to the fidelity between initial and final ground state
and its fidelity susceptibility $\chi_{F}$ \citep{rossini_decoherence_2007}.
Considering the scaling of the fidelity susceptibility $\chi_{F}$
in this regime (see \cite{campos_venuti_quantum_2007}) one obtains
$\max\delta\lambda^{2}\left\{ L^{d},L^{2/\nu}\right\} \ll1$ or $\delta\lambda\ll\min\left\{ L^{-d/2},L^{-1/\nu}\right\} $
where $\delta\lambda=\lambda_{2}-\lambda_{1}$ is the quench amplitude.
As is often the case, the symbol {}``$\ll$'' indicates a conservative
estimates and $\delta\lambda\sim\min\left\{ L^{-d/2},L^{-1/\nu}\right\} $
indicates the region where a crossover takes place. Once the small
quench condition is satisfied, universal behavior in the full time
statistics of observables expectation values is expected in the quasi-critical
region when $\xi^{(2)}\gg L$ where $\xi^{(2)}\simeq\left|\lambda_{2}-\lambda_{c}\right|^{-\nu}$
is the correlation length of the evolution Hamiltonian. Roughly speaking
the condition to observe universal distribution can be written compactly
as $\xi^{(j)}\gg L$ with $j=1,2$ . Moreover $L$ should be large
enough such that i) the universality in the function $Q_{2p}(E)$
sets in and ii) the finite-size corrections to the zeta function results
are small. Both of these conditions depend on the critical exponent
$2\alpha$. For larger values of $2\alpha-d$, a critical, universal
distribution can be observed for smaller sizes. In the opposite, off-critical
region $\xi^{(2)}\ll L$, temporal distributions are expected to be
of Gaussian shape \cite{campos_venuti_universality_2010,campos_venuti_exact_2011}.

We have verified universality in temporal distributions on the hand
of the XY model in transverse field given by the Hamiltonian
\begin{equation}
H=\sum_{j=1}^{L}\left[\frac{1+\gamma}{2}\sigma_{j}^{x}\sigma_{j+1}^{x}+\frac{1-\gamma}{2}\sigma_{j}^{y}\sigma_{j+1}^{y}+h\sigma_{j}^{z}\right],\label{eq:XY}
\end{equation}
defined on a chain of $L$ sites with periodic boundary conditions
for the spins, i.e.~$\sigma_{L+1}^{\alpha}=\sigma_{1}^{\alpha}$,
$\alpha=x,y$. The system is initialized in the ground state of Hamiltonian
(\ref{eq:XY}) with parameters $h=h_{1}$ and $\gamma=\gamma_{1}$
which are then suddenly changed to $h_{2}$ and $\gamma_{2}$. The
transverse magnetization at time $t$ has the form
\begin{align}
\mathcal{M}(t) & =\sum_{j}\langle\sigma_{j}^{z}(t)\rangle\\
 & =\sum_{k}\left[\cos(\vartheta_{k}^{(2)})\cos(\delta\vartheta_{k})+\right.\\
 & \left.\sin(\vartheta_{k}^{(2)})\sin(\delta\vartheta_{k})\cos(t\Lambda_{k})\right]
\end{align}
where $k$ are ABC momenta for the fermions: $k=2\pi/L\left(n+1/2\right)$,
$n=0,1,\ldots,L-1$ and the Bogoliubov angles are given by $\tan\vartheta_{k}^{(i)}=-\gamma^{(i)}\sin(k)/(h^{(i)}+\cos(k))$
and $\delta\vartheta_{k}=\vartheta_{k}^{(2)}-\vartheta_{k}^{(1)}$.

\subsection{Scaling behavior}

Let us write $Q_{p}=Q_{1}^{p}{\mathrm{tr}}\,\Omega^{p},$ where we
have defined $\Omega:=Q_{1}^{-1}{\mathrm{diag}}\,\left(|Z_{n}|\right)_{n>0}.$
Since $\Omega$ is dimensionless the scaling behavior of $Q_{p}$
is clearly dictated by the scaling dimension of $Q_{1}^{p}$ this
latter in turn is just $p$ times the one of $Q_{1}.$ Now $Q_{1}\ge|\tilde{Q}_{1}|$
where $\tilde{Q}_{1}:=\sum_{n>0}Z_{n}=\sum_{n>0}\langle0|A|n\rangle\langle n|B|0\rangle/(E_{n}-E_{0}),$
therefore the scaling dimension $\Delta_{Q_{1}}$ of $Q_{1}$ is lower
bounded by the one of $\tilde{Q}_{1}.$ In formulae $\Delta_{Q_{p}}=\Delta_{Q_{1}^{p}}=p\Delta_{Q_{1}}\ge p\Delta_{\tilde{Q}_{1}}.$
Now we observe that $\tilde{Q}_{1}$ can be written as the time integral
of a connected (imaginary time) two point cross-correlation function
of the observables $A$ and $B:$ 
\begin{eqnarray}
\tilde{Q}_{1} & = & \sum_{n>0}\int_{0}^{\infty}d\tau e^{-\tau(E_{n}-E_{0})}\langle0|A|n\rangle\langle n|B|0\rangle\nonumber \\
 & = & \sum_{n>0}\int_{0}^{\infty}d\tau\langle0|e^{\tau H}Ae^{-\tau H}|n\rangle\langle n|B|0\rangle\nonumber \\
 & = & \int_{0}^{\infty}d\tau\langle A(\tau)B(0)\rangle_{c}
\end{eqnarray}
 where $\langle A(\tau)B(0)\rangle_{c}:=\langle0|A(\tau)B(0)|0\rangle-\langle0|A|0\rangle\langle0|B|0\rangle$.
Assuming now that these operators are local i.e., $X=\sum_{j}X_{j},\,(X=A,B)$
in the continuous limit one finds $\tilde{Q}_{1}\sim\int d^{d}xd^{d}y\int d\tau\, G_{AB}(x,\tau;y,0)$
where $G_{AB}(x,\tau_{1};y,\tau_{2}):=\langle A(x,\tau_{1})B(y,\tau_{2})\rangle_{c}.$
Performing the scaling transformation $x,y,\tau\mapsto\lambda x,\lambda y,\lambda^{\zeta}\tau$
and using the definition of scaling dimension of $A$ and $B$ i.e.,
$G_{AB}\mapsto\lambda^{-\Delta_{A}-\Delta_{B}}G_{AB}$ one finds that
$\Delta_{\tilde{Q}_{1}}=2d+\zeta-\Delta_{A}-\Delta_{B}.$ Finally
assuming that $\Delta_{Q_{1}}=\Delta_{\tilde{Q}_{1}}=:\alpha$ we
recover the key scaling relation used in the main text i.e., $\Delta_{Q_{2p}}=2p\alpha.$

\subsection{Proof of Eq.~(\ref{eq:ratios})}

Let us define the truncated version of $\zeta_{\boldsymbol{b}}\left(\alpha\right)$
as 
\begin{equation}
\zeta_{\boldsymbol{b}}\left(\alpha,L\right):=\sum_{n_{1}=1}^{L}\cdots\sum_{n_{d}=1}^{L}\left\Vert \boldsymbol{n}+\boldsymbol{b}\right\Vert ^{-\alpha}.
\end{equation}
 Note that $\zeta_{\boldsymbol{b}}\left(\alpha,L\right)=\zeta_{\boldsymbol{b}}\left(\alpha\right)-\zeta_{\boldsymbol{b}+\boldsymbol{L}}\left(\alpha\right)$
with $\boldsymbol{L}=\left(L,L,\ldots,L\right)$. Now, for $\alpha\neq d$
one has 
\begin{equation}
\zeta_{\boldsymbol{b}}\left(\alpha,L\right)=\zeta_{\boldsymbol{b}}\left(\alpha\right)+L^{d-\alpha}\left(\frac{C_{\boldsymbol{b}}}{d-\alpha}+O\left(L^{-1}\right)\right)\label{eq:zL-non-d}
\end{equation}
 where $C_{\boldsymbol{b}}$ is a constant independent of $L$. For
$\alpha=d$ the scaling gets modified to 
\begin{equation}
\zeta_{\boldsymbol{b}}\left(d,L\right)=C'_{\boldsymbol{b}}\ln L+O\left(1\right).\label{eq:zL-d}
\end{equation}
 The finite size ratios are given by $R_{2p}=\zeta_{\boldsymbol{b}}\left(2\alpha p,L\right)/\zeta_{\boldsymbol{b}}\left(2\alpha,L\right)^{p}$.
Plugging in Eqns.~(\ref{eq:zL-non-d}) and (\ref{eq:zL-d}) and taking
the limit $L\to\infty$ one recovers Eq.~(\ref{eq:ratios}).

\subsection{Regular points}

Let us then analyze the universal cumulant ratios $R_{p}$ at gapped
region of the phase diagram. Using norm inequalities one can only
show that $R_{2p}\le1$ for all $p$ whereas to prove the central
limit theorem (CLT) one would need $\lim_{L\to\infty}R_{2p}=0$ for
$p\ge2$. Actually, using Lyapunov condition, it suffices to show
that $R_{4}\to0$. Now at regular point of the phase diagram, the
infrared divergence is cured by the gap $E\ge\Delta$. Moreover, quantum
lattice model do not have divergence in the UV as they have a natural
cutoff. For example, a quasi-relativistic, phenomenological one-particle
dispersion often used to model interacting lattice models is given
by $\epsilon_{k}=\sqrt{\sin\left(k\right)^{2}+m^{2}}$. Now, close
but not exactly at the critical point, the contribution to $Q_{p}$
coming from the one-particle excitations is $Q_{p}\sim\sum_{k}\left(\sin\left(k\right)^{2}+m^{2}\right)^{-p\alpha}$
($\zeta=1$ in this case). Since $1/\epsilon_{k}$ is a bounded function
of $k$ we conclude that $Q_{2p}\propto L$ implying the CLT for the
rescaled variable $\mathcal{R}$ as claimed in the main text.

\subsection{CLT: comparison with the equilibrium case}

Let us now compare the origin of the central limit theorem and of
universality for temporal fluctuations and equilibrium fluctuations.
In the equilibrium framework, outside criticality, an extensive observable
can be considered as a sum of weakly dependent random variables. The
CLT arises from the linked cluster expansion \citep{haag_local_1992}
and the fact that, outside criticality the ground state of a local
Hamiltonian is exponentially clustering as proved in \citep{nachtergaele_lieb-robinson_2006}.
This in turn implies that all the cumulants $\kappa_{n}=\sum_{x_{1},\ldots,x_{n}}\langle A(x_{1})A(x_{2})\cdots A(x_{n})\rangle_{c}$
scale with the volume of the system and the rescaled variable $(A-\langle A\rangle)/\Delta A_{\mathrm{eq}}$
becomes Gaussian in the thermodynamic limit. For the time fluctuations
in the dynamical setting instead, the observable $\mathcal{A}$ can
always be considered as a sum of independent random variables as long
as the energies are rationally independent and the quench sufficiently
small. The CLT in this case is the assertion that none of the random
variables dominate. At the critical point, the equilibrium distribution
is determined by the connected moments $\langle A(x_{1})A(x_{2})\cdots A(x_{n})\rangle_{c}$.
These averages depend on the scaling exponent $\Delta_{A}$ which
dictates the long distance asymptotic but also on universal prefactors.
For example for the second cumulant one has $\langle A(x_{1})A(x_{2})\rangle\simeq C_{2}/\left\Vert x_{1}-x_{2}\right\Vert ^{-2\Delta_{A}}$.
Higher cumulants define other universal prefactors which are only
very indirectly related to the exponents $\Delta_{A}$. Instead, in
the dynamical case, there is essentially only one prefactor that becomes
fixed considering the rescaled variable $(\mathcal{A}-\overline{\mathcal{A}})/\Delta\mathcal{A}$.
This ultimately seems to be due to the fact that, for small quenches,
$\mathcal{A}\left(t\right)$ is a sum of independent variables.

\subsection{Proof of Eq.~(\ref{eq:pdf})}

Assuming rational independence of the many-body energies $E_{n}$,
the joint characteristic function can be computed via 
\begin{eqnarray}
\chi\left(\xi,\eta\right) & = & \overline{e^{i\xi X+i\eta Y}}\nonumber \\
 & = & \prod_{n}\int\frac{d\vartheta_{n}}{2\pi}e^{i\xi p_{n}\cos\left(\vartheta_{n}\right)+i\eta p_{n}\sin\left(\vartheta_{n}\right)}\nonumber \\
 & = & \prod_{n}\, J_{0}\left(p_{n}\sqrt{\xi^{2}+\eta^{2}}\right).
\end{eqnarray}
 Let us now compute the cumulative distribution of the LE, 
\begin{eqnarray}
 &  & \mathrm{Prob}\left(\mathcal{L}<r^{2}\right)=\nonumber \\
 & = & \int\frac{d\xi}{\left(2\pi\right)}\frac{d\eta}{\left(2\pi\right)}\int_{x^{2}+y^{2}<r^{2}}dxdye^{-ix\xi-iy\eta}\chi\left(\xi,\eta\right)\nonumber \\
 & = & \int d\xi d\eta\int_{0}^{r}\rho d\rho\int_{0}^{2\pi}d\phi e^{-i\xi\rho\cos\phi-i\eta\rho\sin\phi}\frac{\chi\left(\xi,\eta\right)}{\left(2\pi\right)^{2}}\nonumber \\
 & = & \int d\xi d\eta\int_{0}^{r}\rho d\rho\, J_{0}\left(\rho\sqrt{\xi^{2}+\eta^{2}}\right)\frac{\chi\left(\xi,\eta\right)}{\left(2\pi\right)}\nonumber \\
 & = & \int d\xi d\eta\,\frac{rJ_{1}\left(r\sqrt{\xi^{2}+\eta^{2}}\right)}{\sqrt{\xi^{2}+\eta^{2}}}\frac{\chi\left(\xi,\eta\right)}{\left(2\pi\right)}\nonumber \\
 & = & \int_{0}^{\infty}d\rho\, rJ_{1}\left(r\rho\right)\prod_{n}J_{0}\left(p_{n}\rho\right).
\end{eqnarray}
 Substituting $r^{2}\to r$ and differentiating with respect to $r$
we get a very convenient form of the probability density for the Loschmidt
echo, i.e.~Eq.~(\ref{eq:pdf}) 
\begin{eqnarray}
P_{\mathcal{L}}\left(x\right) & = & \int_{0}^{\infty}K\left(\sqrt{x},\rho\right)J\left(\rho\right)d\rho\label{eq:pdf-1}\\
K\left(\sqrt{x},\rho\right) & = & \frac{J_{1}\left(\sqrt{x}\rho\right)}{2\sqrt{x}}+\frac{\rho}{4}\left[J_{0}\left(\sqrt{x}\rho\right)-J_{2}\left(\sqrt{x}\rho\right)\right]\nonumber \\
J^{\mathcal{L}}\left(\rho\right) & = & \prod_{n}J_{0}\left(p_{n}\rho\right).\nonumber 
\end{eqnarray}

\subsection{Quasi-free systems}

The formalism developed in the main text does not directly apply to
quasi-free systems because the many-body energies are massively rationally
dependent. In this case the analysis has been carried out in \citep{campos_venuti_exact_2011}.
If the initial state has covariance matrix $R_{x,y}=\langle c_{y}^{\dagger}c_{x}\rangle$,
and the Hamiltonian is $H=\sum_{xy}c_{x}^{\dagger}M_{x,y}c_{y}$,
the LE can be expressed as $\mathcal{L}\left(t\right)=\left|\det\left(1-R+Re^{-itM}\right)\right|^{2}$.
Now, if $\left[M,R\right]=0$ as it happens for quenches, both matrices
can be diagonalized simultaneously and one gets 
\begin{equation}
\mathcal{L}\left(t\right)=\prod_{k}\left(1-\alpha_{k}\sin^{2}\left(t\epsilon_{k}/2\right)\right)
\end{equation}
 where $\epsilon_{k}$ are the one-particle energies, and $\alpha_{k}=4r_{k}\left(1-r_{k}\right)$
and $r_{k}$ are the eigenvalues of $R$. In this case it's easier
to get the probability distribution for the logarithmic LE $\mathcal{G}=\ln\mathcal{L}$.
Assuming now rational independence for the \emph{one-particle energies},
we can obtain its characteristic function 
\begin{equation}
\overline{e^{i\lambda\mathcal{G}}}=\prod_{k}{}_{2}F_{1}\left(\frac{1}{2},-i\lambda,1,\alpha_{k}\right).
\end{equation}
 Now for any finite quench, all the cumulants of $\mathcal{G}$ are
extensive so that the rescaled variable $(\mathcal{G}-\overline{\mathcal{G}})/\Delta\mathcal{G}$
tends in distribution to a Gaussian in the thermodynamic limit \citep{campos_venuti_exact_2011}.
In the critical, small quench scenario, a good approximation to the
distribution of $\mathcal{G}$ is obtained retaining few (e.g.~2)
lowest weights $\alpha_{k}$ and it acquires a double peaked form
as shown in \citep{campos_venuti_exact_2011}.
\end{document}